\documentclass[preprint]{revtex4-1}
	\usepackage{graphicx}
	\begin{document}
		\title{Low-temperature thermal conductivity\\ in polycrystalline graphene}
	\author{D.V. Kolesnikov, V.A. Osipov}
	\affiliation{ Bogoliubov Laboratory of Theoretical Physics,\\ Joint
	Institute for Nuclear Research,\\ 141980 Dubna, Moscow region, Russia }
\date{\today}


	
	\begin{abstract}
	The low-temperature thermal conductivity in polycrystalline graphene
	is theoretically studied. The contributions from three branches of
	acoustic phonons are calculated by taking into account scattering on
	sample borders, point defects and grain boundaries. Phonon
	scattering due to sample borders and grain boundaries is shown to
	result in a $T^{\alpha}$-behaviour in the thermal conductivity where
	$\alpha$ varies between 1 and 2. This behaviour is found to be more
	pronounced for nanosized grain boundaries.
	\\
	{65.80.Ck} {Thermal properties of graphene}\\
	{81.05.ue} {Graphene}\\
	{73.43.Cd} {Theory and modeling }
	\end{abstract}

\maketitle

	\section{Introduction}
	
	The thermal conductivity of both single-layer and few-layer graphene
	is of current experimental and theoretical interest (see a recent
	review~\cite{nika} and the references therein). It was shown that
	the heat conduction in graphene has a specific behaviour due to the
	unique nature of two-dimensional phonons. Main attention has been
	attracted to studies of the heat transfer in graphene near room
	temperature because of possible applications in electronics and
	photonics.
	
	In this paper, we focus on the thermal conductivity in graphene at
	low temperatures. Our interest is stimulated by recent observations
	that the large-scale graphene films are typically
	polycrystalline~\cite{li,yazyev,yazyev2,dasilva} and consist of many
	single-crystalline grains separated by grain boundaries
	(GB)~\cite{huang,lahiri,kim}. The grain sizes are dependent on
	growth conditions, ranging from hundreds of nanometres to tens of
	micrometres for slight changes in growth conditions~\cite{huang}.
	Recently, the fabrication of nanocrystalline graphene made by
	electron-radiation induced cross-linking of aromatic self-assembled
	monolayers and their subsequent annealing was
	reported~\cite{turchanin}. Nanosized grain boundary loops
	originating from paired five- and seven-membered ring disclinations
	were observed and analysed in~\cite{cockayne}.
	
	The GB-induced phonon scattering has a marked impact on the thermal
	conductivity, $\kappa$, at low temperatures. In bulk materials, the
	well-pronounced crossover from $\kappa\sim T^2$ to $\kappa\sim T^3$
	has been proved with the crossover temperature strongly depending on
	the GB size~\cite{osipov,krasavin}. Therefore, one can expect a
	similar effect in two-dimensional polycrystalline materials like
	graphene. The paper is organized as follows. In Section 2 we
	overview the general formalism with three main acoustic phonon
	branches taken into account: longitudinal (LA), transverse (TA) and
	flexural (ZA). The effective temperatures characterizing the phonon
	mean free paths for main scattering mechanisms are introduced. The
	results of numerical calculations of the thermal conductivity are
	presented in section 3. Conclusion is devoted to the discussion of
	the results obtained.
	
	\section{General formalism}
	
\begin{figure}
	\includegraphics[scale=0.35]{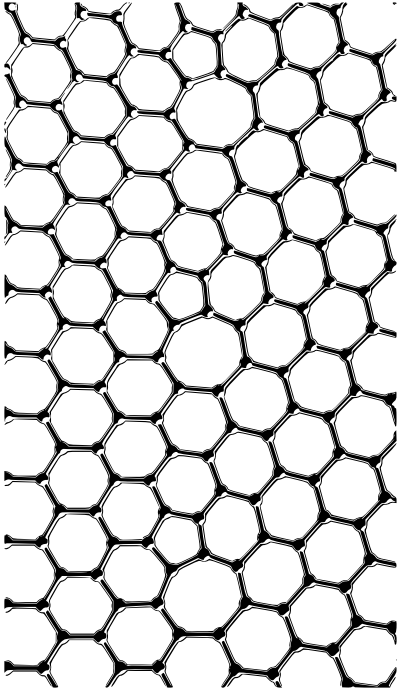} \caption{The fragment of grain
	boundary in polycrystalline graphene. The size $\cal L$ of this wall
	exceeds eight lattice translations. Notice that the whole wall
	corresponds to a single BWDD. }\label{f.0}
	\end{figure}

	According to ~\cite{balandin}, the phonon thermal conductivity of
	single-layer graphene can be written as
	\begin{eqnarray}
	\kappa = \frac{1}{4\pi k_B T^2 h_{eff}}\sum_s \int_0^{q_{D}}(\hbar \omega_s(q))^2 v_s(q) l_s(q,T) \times\nonumber\\
	\times\frac{e^{\hbar\omega_s(q)/(k_B T)}}{(e^{\hbar\omega_s(q)/(k_B T)}-1)^2}q dq,
	\label{balandin}
	\end{eqnarray}
	where summation is performed over the phonon polarization branches
	with the dispersion relations $\omega_s(q)$ ($q$ is the wavevector,
	$q_{D}$ corresponds to the edge of the Brillouin zone, and $s$ is
	the branch index), $h_{eff}$ is the effective graphene layer
	thickness, $v_s(q)=\partial\omega_s(q)/\partial q$ is the phonon
	group velocity, $l_s(q,T)$ is the phonon mean free path, and $k_B$
	is the Boltzmann constant. The phonon mean free path, $l_s$, is
	limited by three major independent scattering mechanisms: sample
	borders (rough boundary (RB)), point defect (PD), and grain
	boundary. Within the relaxation-time approximation, it reads
	\begin{equation}
	1/l_s = 1/l_{0} + 1/l_{PD} + 1/l_{GB},
	\end{equation}
	where $l_0,\; l_{PD}$ and  $l_{GB}$ comes from phonon scattering by
	RB, PD, and GB, respectively. The phonon-RB scattering
	is given by~\cite{Ziman}
	\begin{equation}
	l_0^{-1} = \frac{1}{d}\;\frac{1-p}{1+p},\label{lbound}
	\end{equation}
	with $p$ being the specularity parameter defined as a probability of
	specular scattering at the sample borders and $d$ the graphene layer
	size. The phonon-PD scattering can be written as
	\begin{equation}
	l_{PD}^{-1} = \frac{S_0 \Gamma}{4} \frac{q(\omega) }{v^2(\omega)}\omega^2,\label{lpd}
	\end{equation}
	where $S_0$ is the cross-section area per one atom of the lattice
	and $\Gamma$ is the measure of the scattering strength. It was
	suggested in~\cite{Li} that grain boundaries can be naturally
	described by disclinations. In graphene, the alternating
	pentagon-heptagon structure along high-angle GBs was revealed
	in~\cite{kim}. This finding allows us to model the grain boundary by
	biaxial wedge disclination dipole (BWDD) of length $\cal{L}$ which
	consists of 5-7 pairs (see Fig.~\ref{f.0}). In this case, the phonon-GB scattering can be
	expressed as~\cite{osipov}
	\begin{equation}
	l_{GB}^{-1} = 2D^2(\nu {{\cal{L}}})^2\; n_i\; q\; \mathcal{G}( q {\cal{L}}),
	\label{eqGB}
	\end{equation}
	where $\nu$ is the Frank index, $n_i$ is the areal density of BWDDs,
	$$
	\mathcal{G}(z) = J_0^2(z)+J_1^2(z)-J_0(z)J_1(z)/z,
	$$
	$J_i(z)$ is the Bessel functions of i{\it{th}} kind, and
	\begin{equation}
	D = \pi \gamma (1-2\sigma)/(1-\sigma)\label{eqD},
	\end{equation}
	where $\gamma$ is the Gruneisen constant of a given phonon branch
	and $\sigma$ is the Poisson constant (see \cite{osipov,krasavin} for
	detail).
	
	Generally, there are six phonon branches. However, at low
	temperatures only three branches are of importance: acoustic
	longitudinal (LA) and transversal (TA) phonons with the ordinary
	dispersion relation $\omega_s = v_s q$ ($s=LA,TA$) and out-of-plane
	(flexural) acoustic  phonon branch (ZA) with $\omega_{ZA} = q^2/2m$,
	where $m=2\sqrt{\rho_{2D}/K}$ is an effective parameter~\cite{castroneto}, $\rho_{2D}$ and $K$ are graphene 2D mass density and bending stifness, respectively. Notice that at sufficiently
	low temperatures the umklapp processes of phonon-phonon scattering
	are mainly frozen and can be neglected.
	
	Let us introduce the dimensionless parameter $x,\; x=\hbar \omega/(k_B
	T)$. For LA and TA phonons eq.(\ref{balandin}) takes the form
	\begin{equation}
	\kappa_s = \frac{C_s T^2}{h_{eff}}\int_0^{\theta_D/T} \lambda_s(x,T)\frac{x^3 e^x dx}{(e^x-1)^2},\label{ks}
	\end{equation}
	where $\theta_D$ is the Debye temperature, $C_s = k_B^3
	l_0/(4\pi\hbar^2 v_s)$, and
	\begin{equation}
	\lambda_s^{-1}(x,T)=1+ \left (  x \frac{T}{T_{PD}} \right )^3 + x \frac{T}{T_{GB}} \mathcal{G}(x\frac{T}{T_0})\label{lambdas}
	\end{equation}
	is the total mean free path normalized to $l_0$
	($\lambda_s=l_s/l_0$). As a matter of convenience we have introduced
	here three characteristic temperatures:
	\begin{equation}
	T_{PD} = \frac{\hbar v_s }{k_B} \left(\frac{4}{S_0\Gamma
	l_0}\right)^{1/3}   \label{eqtpd}
	\end{equation} for PD scattering,
	\begin{equation}
	T_0 = \frac{\hbar v_s}{{\cal{L}} k_B}\label{eqt0},
	\end{equation}  and
	$ T_{GB} = \hbar v_s/(2 D^2 \nu^2 {\cal{L}}^2 k_B l_0 n_i)$ for GB
	scattering. In order to clarify the role of the GB scattering let us
	temporary disregard the second term in eq.~(\ref{lambdas}) which is
	responsible for PD scattering. In this case, $l_s\approx l_0(1+2
	T_0/(\pi T_{GB}) +\mathcal{O}(T_0^3/(T_{GB} T^2))^{-1}$ for
	$xT>>T_0$ (the short wavelength limit). As is shown below, typically
	$T_0>T_{GB}$. Therefore, namely $T_0$ defines the region of marked
	influence of phonon-GB scattering and can be considered as a
	threshold temperature. It should be also mentioned that $T_0$
	characterizes the temperature when the wavelength of an incident
	phonon becomes comparable with the size of grain boundary.
	
	To take into account the imperfect packing of grains one can
	introduce the average distance between grain boundaries, $a$, so
	that the areal density of grains is estimated as
	$n_i=(a+{\cal{L}})^{-2}$ and
	\begin{equation}
	T_{GB} = \frac{\hbar v_s (1+a/{\cal{L}})^2}{2 D^2 \nu^2  k_B l_0}, \label{eqtGBa}
	\end{equation}
	where $(1+a/{\cal{L}})^{-2}$ is a packing coefficient. Notice that while $T_{GB}$ depends on $a/{\cal{L}}$, $T_0$ is a function of ${\cal{L}}$ only.
	
	The heat conductivity due to ZA phonons is written as
	\begin{equation}
	\kappa_{ZA} = \frac{C_{ZA} T^{3/2}}{h_{eff}}  \int_0^{{\theta_{ZA}}/{T}} 	\lambda_{ZA}(x,T)\frac{x^{5/2} e^x dx}{(e^x-1)^2},\label{kZA}
	\end{equation}
	where $C_{ZA} = \sqrt{2m} {k_B^{5/2} l_0}/(4\pi\hbar^{3/2})$ and
	\begin{eqnarray}
	\lambda_{ZA}^{-1}(x,T)=1+ \gamma^2_{ZA}(x,T) (  (
	x\frac{T}{T_{PD}} )^{3/2}  + \nonumber\\
	+ \sqrt{ x\frac{T}{T'_{GB}} } \mathcal{G}(\sqrt{x\frac{T}{T_0} }) ).
	\end{eqnarray}
	The characteristic temperature of PD scattering takes the form
	\begin{equation}
	T_{PD} = \frac{1}{m}\left(\frac{4\sqrt{2} }{S_0\Gamma l_0}\right)^{2/3} \frac{\hbar}{k_B},
	\end{equation}
	while for GB scattering one obtains $T_0=\hbar/(2{\cal{L}}^2 m k_B)$ and
	\begin{equation}
	T'_{GB} = \frac{\hbar (1+a/{\cal{L}})^4}{8 mk_B [D'^2 \nu^2 l_0]^2},
	\end{equation}
	where $D' = \pi(1-2\sigma)/(1-\sigma)$. Notice that the Gruneisen
	constant for ZA phonons $\gamma(q)$ is assumed to depend on the
	wavevector $q$ in contrast to LA/TA cases. In accordance with
	\cite{gruneisen} one can use an approximate expression $\gamma_{ZA}(q) = -1-80(q/q_D - 1)^2$ or, equivalently,
	\begin{equation}
	\gamma_{ZA}(x,T) = -1-80(\sqrt{x\frac{T}{\theta_{ZA}}}-1 )^2.\label{gammaZA}
	\end{equation}
	
	\section{Results}
	
	The total phonon thermal conductivity of graphene is calculated by
	using eqs. (\ref{ks}) and (\ref{kZA}). The major initial parameters used in the model are
	$d=5{\mu m}$, $p=0.9$, $v_{LA}=21.3{km/s}$, $v_{TA}=13.6{km/s}$ (from~\cite{balandin}), $\gamma_{LA}=1.8$, $\gamma_{TA}=0.75$ (from~\cite{nika}), and $\gamma_{ZA}(x,T)$ determined by eq. (\ref{gammaZA}) (which approximates the dependence of Gruneisen constant for ZA branch on the wavevector), as well as $m=1.0\cdot 10^6{s/m^2}$  are taken from ~\cite{gruneisen}. One should note, that the Gruneisen parameter for the ZA phonon branch is variating within the Brilluin zone by the order of several magnitudes, in contrast with LA and TA cases. To take into account the influence of ZA phonons, we use eq. (\ref{gammaZA}) for the Gruneisen parameter instead of constant value. We use the upper limit for the specular graphene thickness parameter $h_{eff} = 0.69{nm}$ from~\cite{aabalandin}. One should also note that the differences in the
	recent theoretical and experimental models give rise to the discrepancy in
	the determination of certain graphene constants (see the discussion
	in~\cite{nika} for detail). However, this discrepancy will not play
	crucial role for the effects discussed below.
	
	Using the constants defined above, one obtains  $C_{LA}=8.40\cdot
	10^{-11}{W/K^3}$, $C_{TA}=1.31 \cdot 10^{-10}{W/K^3}$, $C_{ZA} =
	5.93 \cdot 10^{-9}{W/K^{5/2}}$.  The characteristic temperatures are
	given in tables~\ref{t.1} and \ref{t.2}. We restrict our consideration to
	the temperatures below 100 K. To estimate the parameter $T_{PD}$ we fit our
	model in the case of graphene single crystal according to the results
	of~\cite{balandin}. We found that $T_{PD}$ exceeds 100${K}$ for the
	leading phonon modes, and thus PD scattering turns out to be of little
	consequence within the temperature interval considered.
	
\begin{figure*}
	\begin{center}
	\includegraphics[height=6cm]{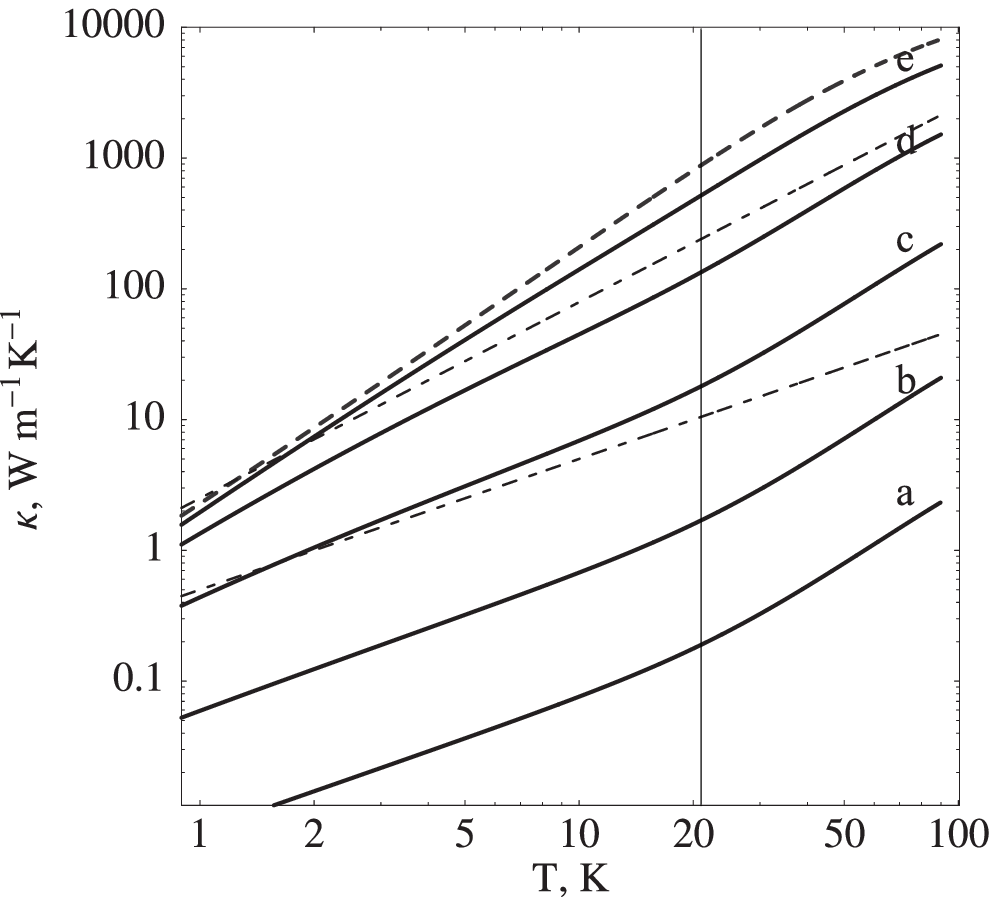}
	\includegraphics[height=6cm]{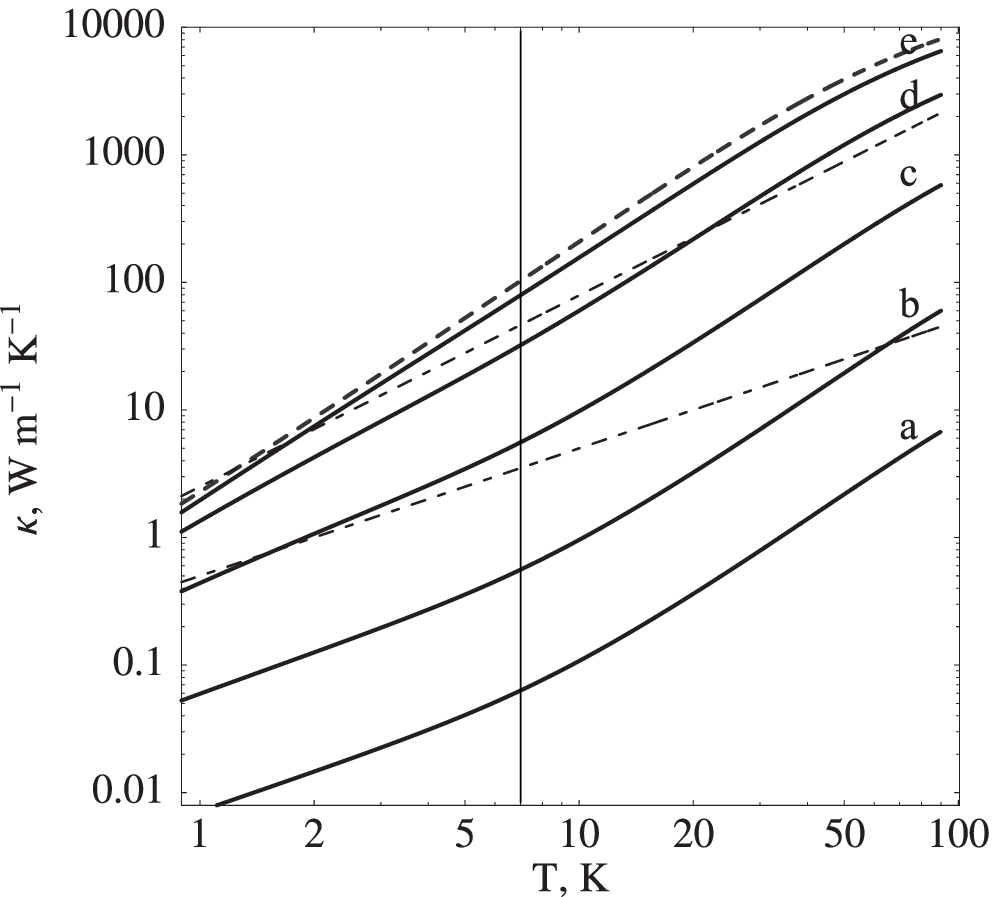}
	\caption{The total thermal conductivity of polycrystalline graphene
	with high-angle grain boundaries for ${\cal{L}}=2.46{nm}$ (left) and
	${\cal{L}}=7.38{nm}$ (right) at different packing coefficients: (a)
	$a/{\cal{L}}=0$, (b) $a/{\cal{L}}=2$, (c) $a/{\cal{L}}=9$, (d)
	$a/{\cal{L}}=30$, and (e) $a/{\cal{L}}=100$ (see table~\ref{t.2}). Dashed
	curve corresponds to a case of graphene single crystal. Dot-dashed
	straight lines shows $T^{1.5}$ and $T$. The approximate crossover
	temperatures are indicated with vertical lines.}
	\label{f.1}
	\end{center}
	\end{figure*}

	Fig.\ref{f.1} shows the thermal conductivity of graphene as a
	function of temperature for ${\cal{L}}=2.46$ nm and ${\cal{L}}=7.38$ nm at
	different packing.
	For comparison, the dot-dashed lines shows the functions $T^{1.5}$ and
	$T$, while the dashed curve corresponds to a case of graphene single
	crystal. The thermal conductivity of the graphene single crystal follows a
	square law in a wide temperature region excepting the very low
	temperatures where $\kappa_{ZA}$ has a marked impact and high temperatures
	(in the range of
	$80-100\, K$) where PD scattering has a slight influence.
	
	For polycrystalline graphene, one can clearly see the characteristic
	points where crossover occurs. The crossover temperature is estimated as
	$T^{*}\approx
	T_0/2$, where $T_0$ is governed by the dominant phonon mode, (TA).
	Above $T^{*}$ the thermal conductivity behaves like $T^2$ while
	below $T^{*}$ the behaviour is acutely sensitive to $T_{GB}$ (see
	eq.~(\ref{eqtGBa})). In turn, $T_{GB}$ depends distinctly on the
	packing coefficient and GB's angle. Notice that high-angle GBs with
	$\nu\approx 0.083$ were revealed in~\cite{kim}. For this case, the
	typical values of $T_{GB}$ at different packing coefficients are
	shown in table~\ref{t.2}. At low $T_{GB}$ one obtains $\kappa\sim T$ while
	growing $T_{GB}$ results in $\kappa\sim T^{1.5}$ and, finally, in
	$\kappa\sim T^2$. One can conclude that the lower is $T_{GB}$, the more
	pronounced influence of grain boundaries takes place.
	
\begin{figure}
	\includegraphics[scale=0.8]{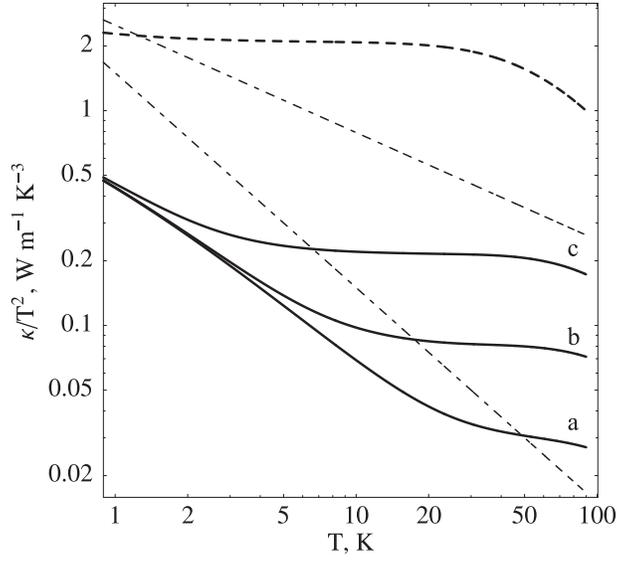}
	\caption{Reduced thermal conductivity, $\kappa/T^2$, versus
	temperature $T$ for a graphene single crystal (dashed) and
	polycrystalline graphene (solid) at fixed packing coefficient
	$a/{\cal{L}}=9$ and different grain sizes: (a) ${\cal{L}}=2.46{nm}$,
	(b) ${\cal{L}}=7.38 {nm}$, and (c) ${\cal{L}}= 22.14 {nm}$.
	Dot-dashed
	straight lines shows $T^{1.5}$ and $T$. }\label{f.2}
	\end{figure}

	Fig. \ref{f.2} shows reduced thermal conductivity, $\kappa/T^2$, versus
	temperature at fixed packing coefficient.  As is clearly seen, both the
	thermal conductivity and the crossover temperature decreases with
	increasing GB size. In case of graphene single crystal, one can mention
	once again a slight deviation from a $T^2$-behaviour due to the influence
	of ZA phonons. Point defects are responsible for deviations at high
	temperatures.
	
	\begin{table}
	\caption{The threshold temperature at different grain
	sizes for three phonon branches.}
	\label{t.1}
	\begin{center}
	\begin{tabular}{lccc}
	\hline ${\cal{L}}$, nm & 2.46 & 7.38 & 22.14 \\
	\hline T$_0$,K (LA) & 66.0 & 22.0 &  7.33 \\
	T$_0$,K (TA) & 42.1 & 14.0 & 4.68  \\
	T$_0$,K (ZA) & 0.6 & 0.066 & 0.0074 \\
	\hline
	\end{tabular}
	\end{center}
	\end{table}
	
	\begin{table}
	\caption{The grain boundary temperature at different effective distances
	between grains for the case of high-angle grain boundaries ($\nu=1/12$).}
	\label{t.2}
	\begin{tabular}{lccccc}
	\hline  $a/{\cal{L}}$ & 0.0 & 2.0 & 9.0 & 30.0 & 100.0\\
	\hline $T_{GB},\, {mK}$ (LA) & 1.5 & 13.5 & 149.7 & 1438.9 & 15273.6 \\
	$T_{GB},\, {mK}$ (TA) & 5.5 & 49.5 & 550.6 & 5291.8 & 56172.5  \\
	$T'_{GB}, {mK}$ (ZA) & $1.3 \cdot 10^{-7}$ & $1.06 \cdot 10^{-5}$  &
	0.0013 & 0.12 & 13.63 \\
	\hline
	\end{tabular}
	\end{table}
	
	\section{Conclusion}
	
	In this paper, the influence of polycrystalline structure on the
	low-temperature heat transfer in suspended graphene has been
	theoretically studied. Three acoustic phonon branches were taken
	into account and three types of scattering mechanisms were
	discussed: sample boundary scattering, point defects and phonon
	scattering on the grain boundaries. The rough-boundary scattering is
	found to play an important role over the whole considered
	temperature range (0-100 K). On the contrary, the point defect
	scattering manifests itself only at the upper limit of the
	temperature interval. The influence of polycrystalline structure is
	described by two temperature parameters $T_{GB}$ and $T_0$ which, in
	turn, define the crossover temperature $T^*$. Above $T^*$, the heat
	conductivity behaves like in graphene single crystal, $\kappa\sim
	T^2$, with an adjusted rough-boundary scattering parameter. Below
	the crossover point we found that $\kappa\sim T^{\alpha}$ where
	$\alpha$ values range from 1 to 2 with increasing $T_{GB}$. In
	experiment, such behaviour of the thermal conductivity could be a
	manifestation of the grain-boundary scattering mechanism. $T_{GB}$
	decreases with an increase of the packing coefficient, the
	misorientation angle and/or a sample size while $T_0$ depends on the
	grain boundary length only.
	
	In our study, the contribution of TA phonons is found to be
	dominant. The contribution of LA phonon branch is suppressed due to
	the higher (in comparison to TA) sound velocity while ZA phonons are
	of little importance due to the large Gruneisen constant. To be more
	precise, in our consideration ZA phonons manifests themselves only
	at very low temperatures and the grain-boundary scattering
	additionally suppresses their influence.
	
	It is interesting to note that in some recent experiments with
	suspended single-layer~\cite{Xu,Xu2} and few-layer~\cite{fewlayer}
	graphene the temperature dependence of $\kappa$  was found to follow
	a power law with an exponent of 1.4$\pm$ 0.1. This behaviour can be
	attributed to a dominant contribution from the ZA phonon modes.
	However, the measured thermal conductivity in those samples are much
	lower compared to the theoretically expected values. As possible
	alternative explanations one considers scattering of phonons in the
	bilayer graphene by a residual polymeric layer, a high concentration
	of defects due to processing~\cite{pettes}, or the polycrystalline
	structure of graphene with small and misoriented
	grains~\cite{aabalandin}. In this article we have shown that the phonon-GB
	scattering will objectively result in $T^\alpha$ behaviour of thermal
	conductivity with $1<\alpha\leq 2$.\\

	This work has been supported by the Russian Foundation for Basic Research
	under grant No. 12-02-01081.

	\end{document}